\documentclass[11pt,a4paper]{article}
\usepackage{jheppub_kim}
\usepackage{rotating} 
 \usepackage{graphicx,epsfig}
\usepackage{amsmath}
\usepackage {amssymb}
\usepackage{subfigure}

\usepackage{relsize}

\usepackage{graphicx}
\usepackage{epsfig}
\usepackage{bm}
\usepackage{amssymb}
\usepackage{float}
\usepackage{amsmath}
\usepackage{dcolumn}
\usepackage{cancel}

\providecommand{\U}[1]{\protect\rule{.1in}{.1in}}

\newcommand{\newc}{\newcommand}

\newc{\be}{\begin{equation}}
\newc{\ee}{\end{equation}}

\newc{\ba}{\begin{eqnarray}}
\newc{\ea}{\end{eqnarray}}
\newc{\bea}{\begin{eqnarray*}}
\newc{\eea}{\end{eqnarray*}}
\newc{\D}{\partial}
\newc{\ie}{{\it i.e.} }
\newc{\eg}{{\it e.g.} }
\newc{\etc}{{\it etc.} }
\newc{\etal}{{\it et al.}}
\newc{\lcdm}{$\Lambda$CDM }

\newc{\ra}{\Rightarrow}

\title{Late-time cosmology with phantom dark-energy in $f(Q)$ gravity}

\author{Andreas Lymperis}

\affiliation{Department of Physics, University of Patras, 26500 Patras, 
Greece}

\emailAdd{alymperis@upatras.gr}

\abstract
{Motivated by the exciting features and a recent proposed general form of the function of non-metricity scalar Q, we investigate the cosmological implications in $f(Q)$ gravity, through the resulting effective dark energy sector, extracting analytical expressions for the dark energy density, equation-of-state and the deceleration parameters. We show that even in the absence of a cosmological constant, the universe exhibits the usual thermal history, with the sequence of matter and dark energy eras, and the dark-energy equation-of-state parameter always lie in the phantom regime. Additionally, calculating the age of the universe, through the extracted analytical equations of the scenario at hand, we show that the result coincide with the value corresponding to $\Lambda$CDM scenario within 1$\sigma$. Moreover, we show the excellent agreement of the scenario at hand with Supernovae type Ia observational data. Lastly, comparing the cosmological behavior in the case of the absence of an explicit cosmological constant, with the one of the presence of a cosmological constant we show that $f(Q)$ gravity can mimic the cosmological constant in a very efficient way, providing very similar behavior, revealing the advantages and capabilitites of the scenario at hand.}

\begin{document}
\maketitle

\section{Introduction}
Undoubtedly, General Relativity (GR) constitutes the pillar in the study of gravitational interactions, and consequently $\Lambda$CDM scenario is its consistent cosmological model. Although the success of GR theory and $\Lambda$CDM model, there are arising problems of theoretical nature such as, the  cosmological constant problem, the coincidence problem the non-renormalizability of GR, etc, and of observational nature such as, the Hubble tension, the $\sigma_8$ tension, etc \cite{Addazi:2021xuf,Abdalla:2022yfr,Martin:2012bt,Freedman:2017yms, Lusso:2019akb,Lin:2019htv,Perivolaropoulos:2021jda}. Thus, the need of extended theories of gravity and models beyond $\Lambda$CDM is more urgent than ever. Hence, various modified theories of gravity have been constructed \cite{CANTATA:2021ktz,Capozziello:2011et,Clifton:2011jh,Cai:2015emx,Nojiri:2017ncd} and several models beyond the $\Lambda$CDM have been proposed \cite{Copeland:2006wr,Joyce:2016vqv,Cai:2009zp} in in the literature in order to give a solution to the aforementioned problems.
            
There are three main directions of gravitational modifications. The first way is by adding extra terms in Einstein-Hilbert Lagrangian, from which way arise modified gravities such as $f(R)$ gravity \cite{Starobinsky:1980te,Capozziello:2002rd,DeFelice:2010aj,Nojiri:2010wj}, Gauss-Bonnet and $f(G)$ gravity \cite{Antoniadis:1993jc,Nojiri:2005jg,DeFelice:2008wz}, cubic and $f(P)$ gravity 
\cite{Erices:2019mkd,Marciu:2020ysf,BeltranJimenez:2020lee}, Horndeski/Galileon scalar-tensor theories \cite{Horndeski:1974wa,Deffayet:2009wt}, etc. The second direction is by adding extra terms to the torsion, namely to the torsional formulation of gravity, resulting to extended theories of gravity, known as modified teleparallel
theories, such as $f(T)$ gravity \cite{Bengochea:2008gz,Cai:2015emx}, $f(T,T_{G})$ gravity \cite{Kofinas:2014owa,Kofinas:2014aka, Kofinas:2014daa}, $f(T,B)$ gravity \cite{Bahamonde:2015zma,Bahamonde:2016grb}, scalar-torsion gravity \cite{Geng:2011aj}, etc. An alternative way and third direction is to use the ``symmetric teleparallel gravity'' as starting point, which is based on the non-metricity scalar $Q$, and construct new extended theories of gravity which contains a function $f(Q)$ in the Lagrangian. \cite{BeltranJimenez:2017tkd}.

This recent class of modification of $f(Q)$ gravity, has attracted a lot of interest in the literature. From its application to cosmology and astrophysics \cite{Delhom:2020vpe,Delhom:2020hkb,Barros:2020bgg,BeltranJimenez:2020sih,Jimenez-Cano:2020chm,DAmbrosio:2020nev,Rubiera-Garcia:2020gcl,Xu:2020yeg,Ayuso:2020dcu,Cabral:2020fax,Flathmann:2020zyj,Frusciante:2021sio,Yang:2021fjy,Fu:2021rgu,Khyllep:2021pcu,Anagnostopoulos:2021ydo,Albuquerque:2022eac,Myrzakulov:2022akb,Chanda:2022cod,Jarv:2018bgs,Hohmann:2017jao,Agrawal:2022vdg,Harko:2021tav,Myrzakulov:2021vel,Iosifidis:2021kqo,Mandal:2021wer}, to interesting cosmological
phenomenology at the background level \cite{Jimenez:2019ovq, Dialektopoulos:2019mtr, Bajardi:2020fxh, Flathmann:2020zyj,Mandal:2020buf,DAmbrosio:2020nev,Mandal:2020lyq, Dimakis:2021gby, Nakayama:2021rda, Khyllep:2021pcu,Hohmann:2021ast,Wang:2021zaz,Quiros:2021eju,Ferreira:2022jcd,Solanki:2022ccf,De:2022shr,Solanki:2021qni,Capozziello:2022wgl,Narawade:2022jeg,Dimakis:2022rkd,Dimakis:2022wkj,Albuquerque:2022eac,Arora:2022mlo,Pati:2022dwl,Khyllep:2022spx}. Additionally, it can fit observations in a very satisfactory way, through its confrontation with various background and perturbation observational data \cite{Soudi:2018dhv,Lazkoz:2019sjl,Barros:2020bgg,Ayuso:2020dcu, Anagnostopoulos:2021ydo,Mandal:2021bpd,Atayde:2021pgb,Frusciante:2021sio}, and moreover, it passes the Big Bang Nucleosynthesis (BBN) constraints too \cite{Anagnostopoulos:2022gej}. All these 
exciting features of  $f(Q)$ gravity reveal that $f(Q)$ gravity may challenge the standard $\Lambda$CDM scenario.

The plan of the manuscript is the following. In Section \ref{sec:f_Q_comology} we briefly review $f(Q)$ gravity and the solutions of the field equations of the theory in the cosmological framework, alongside the  specific  $f(Q)$ model that will be studied. In Section \ref{cosmev} we investigate the cosmological evolution of the universe focusing on the behavior of the dark energy density and equation-of-state parameters, in the cases where a cosmological constant is absent and present. Additionally, we calculate the age of the universe with the scenario at hand and confront its cosmological behavior with Supernovae type Ia (SN Ia) data, in order to present the behavior of the model more transparently. Finally, in Section \ref{concl} we summarize our results.

\section{$f(Q)$ cosmology: a brief overview} \label{sec:f_Q_comology}
The total action of $f(Q)$ gravity is given by \cite{BeltranJimenez:2017tkd,Jimenez:2019ovq}
\be  \label{action00}
S_{tot}=\int \left[ -\frac{1}{2\kappa^{2}}f(Q)+\mathcal{L}_m\right]\sqrt{-g}~d^4x,
\ee
where $\kappa^{2} =8\pi G$ is the gravitational constant, $g$ is the determinant of the metric $g_{\mu\nu}$ and $\mathcal{L}_m$ is the matter Lagrangian density. The non-metricity scalar is defined as  
\cite{BeltranJimenez:2017tkd}  
  \begin{equation}
\label{NontyScalar}
Q=-\frac{1}{4}Q_{\alpha \beta \gamma}Q^{\alpha \beta 
\gamma}+\frac{1}{2}Q_{\alpha \beta \gamma}Q^{ \gamma \beta 
\alpha}+\frac{1}{4}Q_{\alpha}Q^{\alpha}-\frac{1}{2}Q_{\alpha}\tilde{Q}^{\alpha} 
\,,
\end{equation}
where
\be
Q_{\alpha}\equiv Q_{\alpha \ \mu}^{\ \: \mu}, \ \ \ \ \ \ \ \  
\tilde{Q}^{\alpha} 
\equiv Q_{\mu }^{\ \: \mu \alpha}
 \ee
are obtained  from contractions of the  non-metricity tensor $
    Q_{\alpha\mu\nu}\equiv\nabla_\alpha g_{\mu\nu}$. Hence, in the case where $f(Q)=Q$, Symmetric Teleparallel Equivalent of General 
Relativity (STEGR) and thus General Relativity (GR), is recovered.

The field equations are obtained from variation of the action (\ref{action00}), which read as
\cite{Jimenez:2019ovq, Dialektopoulos:2019mtr}: 
\begin{eqnarray}
&&  
\frac{2}{\sqrt{-g}} \nabla_{\alpha}\left\{\sqrt{-g} g_{\beta \nu} f_{Q} 
\left[- \frac{1}{2} L^{\alpha \mu \beta}+ \frac{1}{4} g^{\mu \beta} 
\left(Q^\alpha -  \tilde{Q}^\alpha \right) \right.\right.\nonumber\\
&&\left.\left. \ \ \ \ \ \ \ \ \ \ \ \ \ \ \ \ \ \  \ \ \ \ \ \ \ \ \ \ \ \, 
- \frac{1}{8} 
\left(g^{\alpha \mu} Q^\beta + g^{\alpha \beta} Q^\mu  
\right)\right]\right\}
 \nonumber \\
&& + f_{Q} \left[- \frac{1}{2} L^{\mu \alpha \beta}- \frac{1}{8} \left(g^{\mu 
\alpha} Q^\beta 
+ g^{\mu \beta} Q^\alpha  \right)
 \right. \nonumber\\
&&\left. \ \ \ \ \ \ \ + \frac{1}{4} g^{\alpha 
\beta} \left(Q^\mu -  \tilde{Q}^\mu \right)
\right] Q_{\nu \alpha 
\beta} +\frac{1}{2} \delta_{\nu}^{\mu} f=\kappa^{2} T_{\,\,\,\nu}^{\mu}\,,
\label{eoms}
\end{eqnarray}
where 
$
L^{\alpha}_{\,\,\mu\nu}=\frac{1}{2}Q^{\alpha}_{\,\,\mu\nu}-Q^{\,\,\,\alpha}_{
(\mu\,\,\,\nu)} $ is  the  disformation tensor, $f_{Q}\equiv\partial f/\partial Q$ and $T_{\mu\nu}$ is the usual matter energy-momentum tensor, defined as
\be \label{eq:Ttensor}
T_{\mu\nu} =-\frac{2}{\sqrt{-g}}\frac{\delta(\sqrt{-g}\mathcal{L}_{m})}{\delta g^{\mu\nu}}.
\ee

In the cosmological framework we assume a spatially flat (k=0), homogeneous and isotropic  Friedmann-Lema\^itre-Robertson-Walker (FLRW) metric, which in a Cartesian coordinates is given by 
\be \label{eq:FLRW}
ds^{2}=-dt^{2}+a^{2}(t)(dx^2+dy^2+dz^2)\,,
\ee
where $a(t)$ is the scale factor and $t$ the cosmic time. In the FLRW spacetime the cosmological solutions of the field equations are the two Friedmann equations
\cite{Jimenez:2019ovq}
\begin{eqnarray}
\label{Fr1}
6f_QH^2-\frac12f &=& 8\pi G\rho_m,  \\
\big(12H^2f_{QQ}+f_Q\big)\dot{H}&=&-4\pi G(\rho_m+p_m)\,,
\label{Fr2}
\end{eqnarray}
where $H\equiv\dot{a}/a$ is the Hubble function, dot
denotes derivative with respect to cosmic time $t$ and  $\rho_m$, $p_m$ are the energy density and pressure of the matter perfect fluid respectively. Finally, the full set of cosmological solutions close with the matter conservation equation 
\be \label{conslaw}
\dot{\rho}_m+3H(\rho_m+p_m)=0.
\ee
Let us note here, that in an FRW background
the nonmetricity scalar $Q$ becomes $Q=6H^2$. 
We can rewrite (\ref{Fr1}), (\ref{Fr2}) as
\begin{eqnarray}
\label{FR3}
3H^{2}&=&\kappa^{2}(\rho_{m}+\rho_{de}) \,,  \\
3H^{2}+\dot{H}&=&-\kappa^{2}(p_{m}+p_{de}) \,,
\label{FR4}
\end{eqnarray}
where we have defined the effective dark energy sector, \ie the dark energy density and pressure, as \cite{Anagnostopoulos:2022gej}
\begin{eqnarray}
\label{endende}
\rho_{de}&=&\frac{3}{\kappa^{2}}\left [H^{2}(1-2f_{Q})+\frac{f}{6} \right ] \,, \\
p_{de}&=&-\frac{1}{\kappa^{2}}\left [2\dot{H}(1-f_{Q})+\frac{f}{2}+3H^{2}(1-8f_{QQ}\dot{H}-2f_{Q})\right ].
\label{presde}
\end{eqnarray}

Following \cite{Anagnostopoulos:2021ydo} we adopt the proposed general form of the function $f(Q)$, which has the expression \cite{Anagnostopoulos:2021ydo} 
\be \label{qform} 
f(Q)=Qe^{\lambda \frac{Q_{0}}{Q}},
\ee
where $\lambda$ is a dimensionless free parameter and $Q_{0}=6H^{2}_{0}$ with $H_{0}$ the current value of the Hubble parameter. Note that the parameter $\lambda$ is the only free parameter of the model and for $\lambda =0$, the cosmological equations reduce to the ones of GR. Additionally, the aforementioned model provide very interesting and satisfactory observational fit for $\lambda \neq 0$, even in the case where a cosmological constant is absent \cite{Anagnostopoulos:2021ydo}. At this point, let us mention that although $f(R)$ or $f(T)$ gravity have many advantages and more specifically some of them can fit the observational data in a very efficient way, the newly proposed extended gravity, namely $f(Q)$ gravity with the exponential form \ref{qform}, has a very interesting cosmological behavior. Beyond the fact that it might be preferred  at least by some cosmological datasets, \eg overperforms the concordance model \cite{Anagnostopoulos:2021ydo}, could solve the $H_0$ tension \cite{Anagnostopoulos:2021ydo,Heisenberg:2022gqk} and satisﬁes trivially the BBN constraints \cite{Anagnostopoulos:2022gej}, it does not face the cosmological constant problem, since it does not include a hidden cosmological constant inside the $f(Q)$ form \ref{qform}, since there is not a parameter value limit which recovers $\Lambda$. Additionally it maintains the same number of free parameters with $\Lambda$CDM cosmology, without the need of extra parameters, and moreover in the far past, namely for $z\gg 1$, it has General Relativity as a limit in the absence of a cosmological constant. Thus, with these features at hand, the exponential $f(Q)$ model, in general, may constitute a very promising alternative to the concordance model.

Using the fact that $f_{Q}=e^{\lambda \frac{Q_{0}}{{Q}}}-\lambda \frac{Q_{0}}{{Q}}e^{\lambda \frac{Q_{0}}{{Q}}}$ and $f_{QQ}=\lambda^{2} \frac{Q_{0}^{2}}{{Q}^{3}}e^{\lambda \frac{Q_{0}}{{Q}}}$ and substituting in (\ref{endende}) and (\ref{presde}) we obtain
\begin{eqnarray}
\label{endende1}
\rho_{de}&=&\frac{3}{8\pi G}\left [H^{2}-\left (H^{2}-2\lambda H^{2}_{0}\right )e^{\lambda \frac{H^{2}_{0}}{H^{2}}} \right ] \,, \\
p_{de}&=&-\frac{1}{8\pi G}\left [(2\dot{H}+3H^{2})+\left (-4\lambda^{2}\frac{H^{4}_{0}}{H^{4}}+2\lambda \frac{H^{2}_{0}}{H^{2}}+2\lambda H^{2}_{0}-H^{2}-2\right )e^{\lambda \frac{H^{2}_{0}}{H^{2}}} \right].
\label{presde1}
\end{eqnarray}
Thus, we can define the equation-of-state parameter for the effective dark energy sector as
\be \label{wde}
w_{de}\equiv \frac{p_{de}}{\rho_{de}}=-1-\frac{2\dot{H}+\left (-4\lambda^{2}\frac{H^{4}_{0}}{H^{4}}+2\lambda \frac{H^{2}_{0}}{H^{2}}+2\lambda H^{2}_{0}-H^{2}-2\right )e^{\lambda \frac{H^{2}_{0}}{H^{2}}}}{3H^{2}-3\left (H^{2}-2\lambda H^{2}_{0}\right )e^{\lambda \frac{H^{2}_{0}}{H^{2}}}}.
\ee

In summary equations (\ref{conslaw}), (\ref{FR3}) and (\ref{FR4}) can determine the universe evolution, as long as the matter equation-of-state parameter is known. In the next section we will investigate this cosmological behaviour through the obtained cosmological equations of the model.

\section{Cosmological evolution} \label{cosmev}
In this section we desire to investigate the cosmological evolution of the universe through modified scenario of $f(Q)$ gravity, namely through the cosmological equations (\ref{FR3}) and (\ref{FR4}). In order to provide analytical solutions of the dark sector, we assume that universe is filled with dust matter, namely $w_{m}=0$ ($p_{m}=0$). In this case the matter conservation equation (\ref{conslaw}) gives $\rho_{m} = \frac{\rho_{m0}}{a^3}$, where in what follows, the subscript ``0" will denote the present value of the corresponding quantity, at the current scale factor which is set to $a_0=1$.

Additionally, introducing the dimensionless parameters 
\begin{eqnarray} \label{omatter}
&&\Omega_m=\frac{8\pi G}{3H^2} \rho_m\\
&& \label{ode}
\Omega_{de}=\frac{8\pi G}{3H^2} \rho_{de},
 \end{eqnarray} 
for the matter and dark energy density sector respectively, the Hubble parameter can be written as
\be \label{h2}
H=\frac{\sqrt{\Omega_{m0}} H_{0}}{\sqrt{a^3 (1-\Omega_{de})}}.
\ee
In what follows we will use the redshift $z$ as the independent variable 
($1+z=1/a$ for $a_0=1$). Hence, differentiation of (\ref{h2}) immediately gives
\be \label{hddot}
\dot H=-\frac{H^2}{2(1-\Omega_{de})}[3(1-\Omega_{de})+(1+z)\Omega'_{de}],
\ee
where prime denotes derivative with respect to the redishift parameter $z$. In the following subsections we investigate the cosmological evolution of the universe, in the case where an explicit cosmological constant is absent and when is present.

\subsection{$\Lambda=0$ case}
In the case where an explicit cosmological constant is absent the cosmological equations are given by (\ref{FR3}) and (\ref{FR4}), where $\rho_{de}$ and $p_{de}$ are given by (\ref{endende1}) and (\ref{presde1}) respectively. Furthermore, in order to provide analytical solutions, it proves convenient to perform  Taylor expansion of the exponential function $e^{\lambda(Q_{0}/Q)}$ around $\lambda=0$, and thus, by using the expression $e^x =1+x+\frac{x^2}{2}+\dots$, and (\ref{h2}), (\ref{endende1}), expansion of the first Friedmann equation (\ref{FR3}) gives
\begin{eqnarray} \label{eqomegade}
&&\left [\frac{\lambda}{\Omega_{m0}(1+z)^{3}}+1 \right 
][1-\Omega_{de}(z)]
\nonumber\\
&&
+\frac{\lambda^{2}}{\Omega^{2}_{m0}(1+z)^{6}}
[1-\Omega_{de}(z)]^{2}\nonumber\\
&&
+\frac{2\lambda^{3}}{\Omega^{4}_{m0}(1+z)^{12
}}[1-\Omega_{de}(z)]^{4}\approx1.
\end{eqnarray}
Additionally, at present time, namely $z=0$, equation (\ref{eqomegade}) immediately gives
\begin{eqnarray}\label{rell0} 
\lambda +\lambda^{2} + 2\lambda^{3} =1-\Omega_{m0}.
\end{eqnarray}
We mention here that the case where $\lambda=0$ corresponds to CDM scenario and in that case leads to dark energy absence (in the absence of a cosmological constant, the scenario at hand does not have $\Lambda$CDM cosmology as a limit). The relation (\ref{rell0}) provides the solution for $\lambda$, namely
\begin{eqnarray} \label{rell00} 
\lambda =-&\frac{1}{6}&-\frac{5}{6\left (62-54\Omega_{m0}+3\sqrt{3}\sqrt{147-248\Omega_{m0}+108\Omega_{m0}^{2}}\right )^{\frac{1}{3}}}+ \nonumber \\
&\frac{1}{6}&\left (62-54\Omega_{m0}+3\sqrt{3}\sqrt{147-248\Omega_{m0}+108\Omega_{m0}^{2}}\right )^{\frac{1}{3}}
\end{eqnarray}
from which $\lambda$ can be eliminated in terms of $\Omega_{m0}$, leaving the scenario at hand with the same number of free parameters with $\Lambda$CDM (note that we have ignored the contribution of radiation). 

Substituting (\ref{rell0}) into (\ref{eqomegade}) we obtain the solutions 
for $\Omega_{de}(z)$, which  read as
 \begin{eqnarray} \label{omegade}
\Omega_{de}= 1&&+\frac{\epsilon_{1}}{2}\left [-\frac{1}{3\mathcal{A}}+\frac{(\lambda -24)}{3\cdot 2^{2/3}\mathcal{A}(\mathcal{B}+\mathcal{C})^{1/3}}+\frac{(\mathcal{B}+\mathcal{C})^{1/3}}{6\cdot 2^{1/3}\mathcal{A}^{2}}\right ]^{1/2} \nonumber \\ 
&&
-\frac{\epsilon_{2}}{2}\left\{-\frac{2}{3\mathcal{A}}-\frac{(\lambda -24)}{3\cdot 2^{2/3}\mathcal{A}(\mathcal{B}+\mathcal{C})^{1/3}}-\frac{(\mathcal{B}+\mathcal{C})^{1/3}}{6\cdot 2^{1/3}\mathcal{A}^{2}} \right. \nonumber \\ 
&&
\left.
+\frac{\left [1+\lambda \left (\frac{\lambda}{\mathcal{A}}\right )^{-1/2}\right ]}{\lambda^{2}\mathcal{A}^{2}\left [-\frac{1}{3\mathcal{A}}+\frac{(\lambda -24)}{3\cdot 2^{2/3}\mathcal{A}(\mathcal{B}+\mathcal{C})^{1/3}}+\frac{(\mathcal{B}+\mathcal{C})^{1/3}}{6\cdot 2^{1/3}\mathcal{A}^{2}}+\mathcal{A}\mathcal{C}^{1/3}\right ]^{1/2}}\right\}^{1/2} \,,
\end{eqnarray} 
 with 
\begin{eqnarray}
\nonumber &&
\!\!\!\!\!\!\!\!\!\!\!\!\!\!\!\!\!\!
\mathcal{A}=\frac{\lambda}{\Omega^{2}_{m0}(1+z)^{6}}, \\ \nonumber
&& \!\!\!\!\!\!\!\!\!\!\!\!\!\!\!\!\!\!\!\!
\mathcal{B}=144\lambda^{2}+2\lambda^{3}+54\left [\left (\frac{\lambda}{\mathcal{A}}\right )^{1/2}+\lambda \right ]^{2}, \\ \nonumber
&& \!\!\!\!\!\!\!\!\!\!\!\!\!\!\!\!\!\!\!\!
\mathcal{C}=\left [-4\left (24\lambda +\lambda^{2}\right )+\mathcal{B}^{2} \right ]^{1/2}.
\end{eqnarray}
and where $\epsilon_{1}, \epsilon_{2}=\pm1$. 
\begin{figure}[!h]
\centering
\includegraphics[width=7cm]{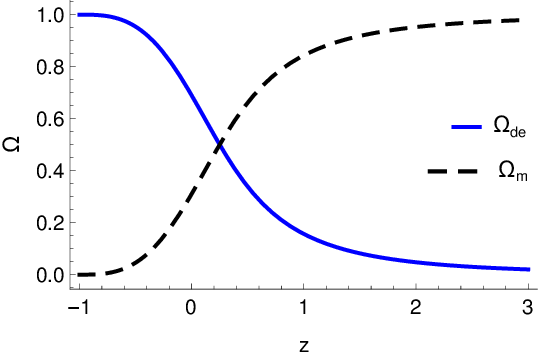}
\includegraphics[width=7cm]{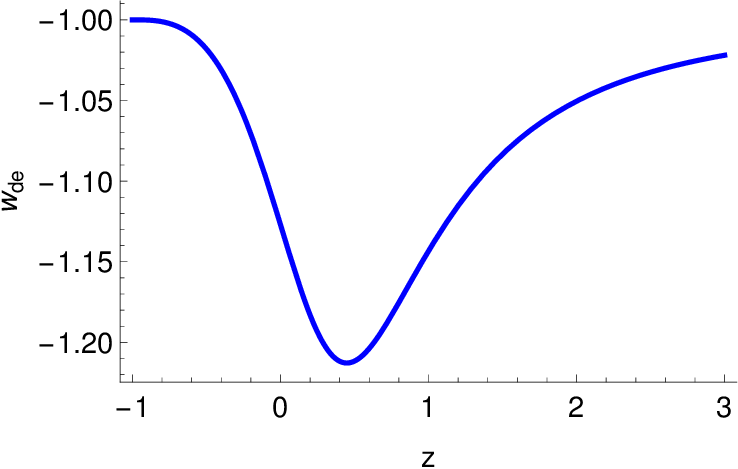}
\includegraphics[width=7cm]{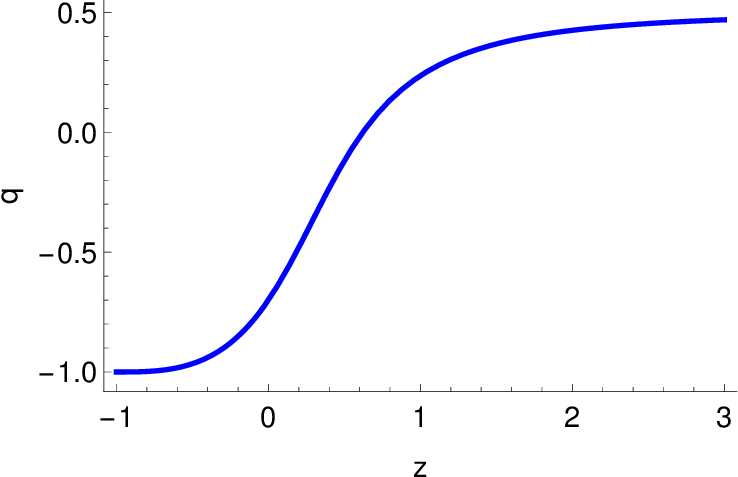}
\caption{\it{{\bf{Upper   graph}}: 
Depiction of the evolution of the  effective dark sector.  {\bf{ Upper graph}}: The
energy density parameter $\Omega_{de}$ (blue-solid) and the
matter density parameter $\Omega_{m}$ (black-dashed) respectively, as a 
function of the redshift $z$, for the
$f(Q)$ gravity scenario, in the case of $\Lambda=0$.
{\bf{ Middle graph}}: The effective dark energy equation-of-state 
parameter $w_{de}$. {\bf{ Lower graph}}: The 
deceleration parameter $q$. In all graphs the parameter, we have imposed $ \Omega_{m0} \approx0.31$ and according to (\ref{rell00}) $\lambda=0.4$ in units of $H_{0}$.}}
\label{OmegasL0}
\end{figure}
As stated above, $\lambda$ can be eliminated from (\ref{rell00}), leaving the scenario at hand with the same degrees of freedom with $\Lambda$CDM scenario, which is an advantage of the model.
With the solution (\ref{omegade}) of the dark energy density at hand,  we can obtain the analytical expression for the  dark energy equation-of-state parameter 
$w_{de}(z)$, by eliminating $\dot{H}$ from (\ref{hddot}) and using (\ref{h2}) and the derivative of (\ref{omegade}). Finally, the other physically interesting quantity, namely the deceleration parameter $q\equiv -1-\frac{\dot H}{H^{2}}$ can be similarly 
extracted by using (\ref{h2}), (\ref{hddot}) and the solution (\ref{omegade}). Let us mention here that from the 4-degree equation (\ref{eqomegade}) we acquire 4 solutions of $\Omega_{de}$, From the four solutions we choose only the solution with
sign (-, +), while we discard all the others, since they lead either to early-time dark energy, or to not physically accepted $\Omega_{de}$, or to the wrong sequence of matter and dark energy epochs.

In Fig. \ref{OmegasL0} we depict the evolution
of the of the dark sector for the scenario at hand. The upper graph shows the evolution of the energy densities $\Omega_{de}$ and $\Omega_{m}=1-\Omega_{de}$. In the upper graph of Fig. \ref{OmegasL0}  we present the evolution of the energy densities $\Omega_{de}$ and $\Omega_{m}$, imposing $\Omega_{m} (z = 0) = \Omega_{m0} = 0.31$ in agreement with the Planck results \cite{Planck:2018vyg}, which corresponds to $\lambda=0.4$ according to (\ref{rell00}). As we can see, we obtain the usual thermal history of the universe, with the sequence of matter and dark-energy epochs, while in the future the universe tends
asymptotically in a dark-energy dominated, de Sitter phase. The middle graph depicts the evolution of $w_{de}$ according to (\ref{wde}), where (\ref{h2}) and (\ref{omegade}) have been used. Finally, in lower graph we show the evolution of the deceleration parameter and as we can see, the transition from deceleration to acceleration takes place at a redshift $z_{tr}\approx0.62$, in agreement with observations. 

\begin{figure}[!h]
\centering
\includegraphics[width=7cm]{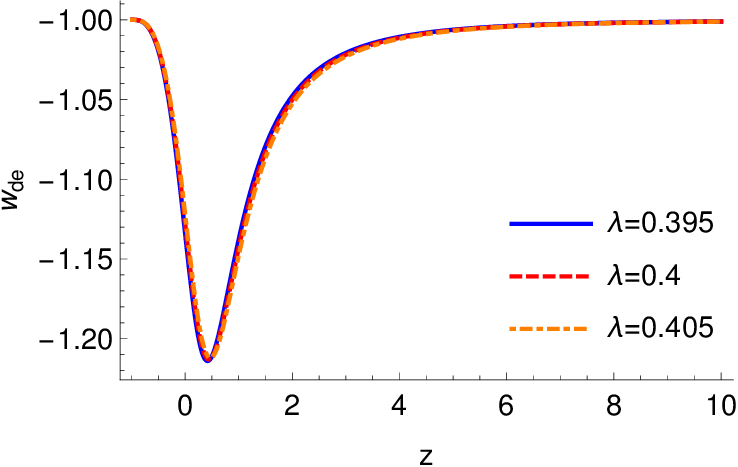}
\caption{\it{The evolution of the equation-of-state parameter $w_{de}$ as a function of 
the redshift parameter $z$, in the case $\Lambda=0$ and for various values of the dimensionless parameter $\lambda$. In all cases we have 
obtained density parameters evolution similar to the graphs of Fig.\ref{OmegasL0}, and $\Omega_{m0}$ lies inside the $2\sigma$ region according to 
Planck Collaboration,  namely $ \Omega_{m0}\approx0.31\pm0.014$ \cite{Planck:2018vyg}}}
\label{multiwdeL0}
\end{figure}

Let us now examine in more detail the role of the parameter $\lambda$ in the evolution, and in particular on $w_{de}$. In Fig. \ref{multiwdeL0} we depict $w_{de}$ 
for different values of $\lambda$. We observe that at high reshifts it slightly lies in the phantom regime, and tends approximatelly to $-1$ (in particular, for $z=10$ we obtain $w_{de}=-1.001$). 
As $z$ decreases to lower values and in particular to present time, the parameter $\lambda$ starts to take effect, deviating from $\Lambda$CDM cosmology, and finally  at asymptotically large times the universe results in a dark-energy dominated, de-Sitter phase. Finally, let us mention that in order to have $ 
\Omega_{m0}\approx0.31\pm0.014$, which is the $2\sigma$ region according to 
Planck Collaboration \cite{Planck:2018vyg}, $\lambda$ is varied in the range 
$0.395\lesssim \lambda \lesssim  0.405$

We see that even in the absence of an explicit cosmological constant, the scenario at hand, is very efficient in mimicking the cosmological constant, providing a very interesting cosmological behavior, despite the fact that in the case where $\Lambda =0$, the exact $\Lambda$CDM cosmology cannot be obtained for any parameter values. This is an advantage of the model alongside with the other capabilities of the scenario at hand.

\subsubsection{Age of the universe}
In order to reveal the capabilities of the scenario at hand in a more thorough way we proceed with the calculation of the universe age. Using the expression
\be
t(z)=\int^{\infty}_{z}\frac{dz^{'}}{(1+z^{'})H(z^{'})}
\ee
and substituting the evolution of $H(z)$ obtained above, we find 
$t(z)=\frac{0.978747}{H_{0}}$. Hence, with $H_0 = (67.27\pm 0.60)$ km/s/Mpc we finally obtain
\be 
t_{age}=13.935^{+0.017}_{-0.017}\ Gyrs,
\ee
which coincides with the value corresponding to 
$\Lambda$CDM scenario, namely  $13.787^{+0.020}_{-0.020}$ Gyrs, within 1$\sigma$.
\cite{Planck:2018vyg}.
In order to show this cosmological behavior in an even more trasparently way, in the next subsection we confront the scenario at hand  with Supernovae type Ia (SN Ia).

\subsubsection{SNIa observational confrontation} \label{sniaconf}
\begin{figure}[ht]
\centering
\includegraphics[scale=0.64]{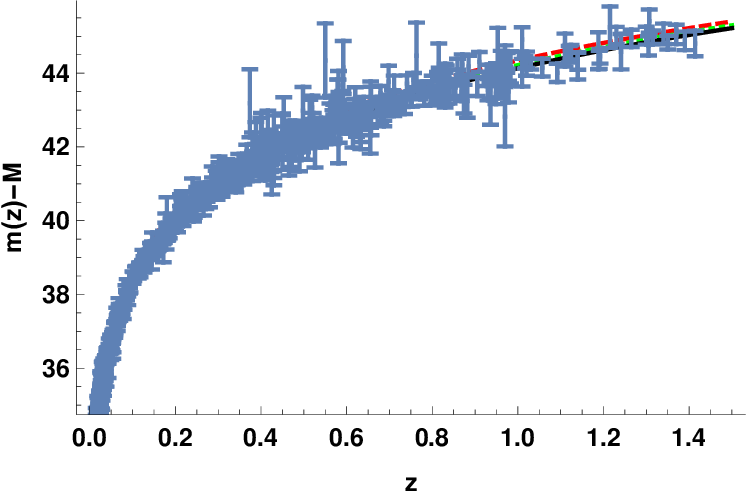}
\caption{
{\it{The theoretically predicted apparent minus absolute magnitude as a function of the 
redshift, for the $f(Q)$ gravity scenario
for $\Lambda=0$, for 
$\lambda=0.3$ (red-dashed) and 
$\lambda=0.4$ (green-dotted) and for $\Lambda$CDM cosmology (black-solid). The observational points correspond to the  $580$ SN Ia data points from \cite{SupernovaCosmologyProject:2011ycw}}}}
\label{SNd}
\end{figure}

The scenario of $f(Q)$ cosmology we study, even 
in the case where an explicit cosmological constant is absent, is efficient in describing the cosmological behavior of the universe. This behavior can present more transparently by confrontation of the scenario at hand  with Supernovae type Ia (SN Ia) data. In particular, the apparent luminosity  $l(z)$, or equivalently the apparent 
magnitude $m(z)$ are related to the luminosity distance through the relation
\be 
2.5 \log\left[\frac{L}{l(z)}\right] = \mu \equiv m(z) - M = 5 
\log\left[\frac{d_L(z)_{\text{obs}}}{Mpc}\right]  + 25,
\ee
where $L$ and $M$ are the absolute luminosity and absolute magnitude respectively.
Additionally, the theoretical value of theluminosity distance is
\be
d_{L}\left(z\right)_\text{th}\equiv\left(1+z\right)
\int^{z}_{0}\frac{dz'}{H\left(z'\right)}~.
\ee
In the scenario at hand, $H(z)$ can be immediately calculated analytically from 
(\ref{h2}), using the solution (\ref{omegade}). Fig. \ref{SNd} depicts the theoretically predicted apparent minus absolute 
magnitude  as a function of $z$, as well as the prediction of 
$\Lambda$CDM cosmology, on top of the   $580$ SN Ia observational data points from 
\cite{SupernovaCosmologyProject:2011ycw}. As we can see the agreement with the SN Ia data is excellent, and the aforementioned $f(Q)$ gravity scenario 
have a slightly higher accelerating behavior.

In summary, this cosmological behavior, shows the capabilities and the advantages of the of the scenario at hand. In the next subsection we desire to confront this case with the case where an explicit cosmological constant is present, in order to show these advantages in a more thorough way.

\subsection{$\Lambda \neq 0$ case}
Let us proceed now in the investigation of the cosmological evolution, when an explicit cosmological constant is present, in order to compare it with the previous case of the absence of a cosmological constant. In this case the two Friedmann equations are   
\begin{eqnarray}
\label{Fr1L}
6f_QH^2-\frac{1}{2}f -\frac{\Lambda}{3} &=& 8\pi G\rho_m \,,  \\
\left (12H^2f_{QQ}+f_Q\right)\dot{H}&=&-4\pi G(\rho_m+p_m)\,,
\label{Fr2L}
\end{eqnarray}
where $\Lambda$ is the cosmological constant. Thus, the cosmological equations are still given by (\ref{FR3}) and (\ref{FR4}), where $\rho_{de}$ and $p_{de}$ can now be written as
\begin{eqnarray}
\label{endende1L}
\rho_{de}&=&\frac{3}{8\pi G}\left [\frac{\Lambda}{3}+H^{2}-\left (H^{2}-2\lambda H^{2}_{0}\right )e^{\lambda \frac{H^{2}_{0}}{H^{2}}} \right ] \,, \\
p_{de}&=&-\frac{1}{8\pi G}\left [\Lambda +(2\dot{H}+3H^{2})+\left (-4\lambda^{2}\frac{H^{4}_{0}}{H^{4}}+2\lambda \frac{H^{2}_{0}}{H^{2}}+2\lambda H^{2}_{0}-H^{2}-2\right )e^{\lambda \frac{H^{2}_{0}}{H^{2}}} \right].
\label{presde1L}
\end{eqnarray}
respectively. Expansion of the first Friedmann equation now gives
\begin{eqnarray} \label{eqomegadeL}
&&\left [\frac{\Lambda}{3H^{2}_{0}\Omega_{m0}(1+z)^{3}}+\frac{\lambda}{\Omega_{m0}(1+z)^{3}}+1 \right 
][1-\Omega_{de}(z)]
\nonumber\\
&&
+\frac{\lambda^{2}}{\Omega^{2}_{m0}(1+z)^{6}}
[1-\Omega_{de}(z)]^{2}\nonumber\\
&&
+\frac{2\lambda^{3}}{\Omega^{4}_{m0}(1+z)^{12
}}[1-\Omega_{de}(z)]^{4}\approx1 ,
\end{eqnarray}
and application of (\ref{eqomegadeL}) at present immediately gives
\be \label{relL}
\Lambda =3H^{2}_{0}(1-\Omega_{m0})-3H^{2}_{0}\left(\lambda +\lambda^{2}+2\lambda^{3}\right),
\ee
leaving the scenario with one free parameter, as one can
eliminate one of the two parameters in terms of the observationally determined quantities $\Omega_{m0}$ and $H_0$.
\begin{figure}[!h]
\centering
\includegraphics[width=7cm]{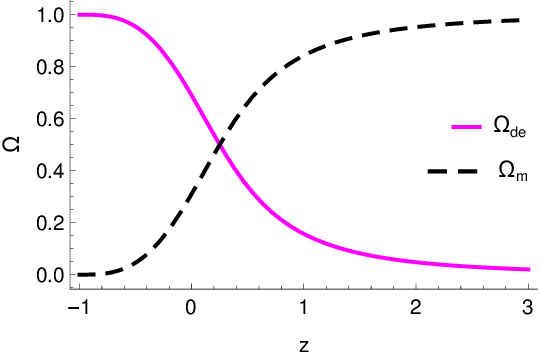}
\includegraphics[width=7cm]{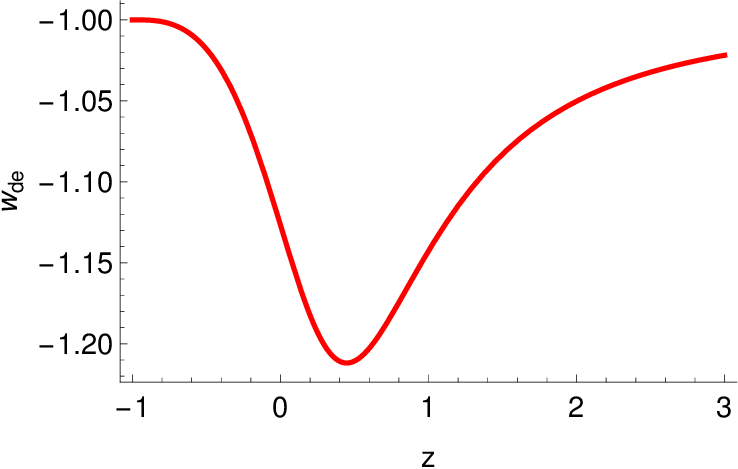}
\includegraphics[width=7cm]{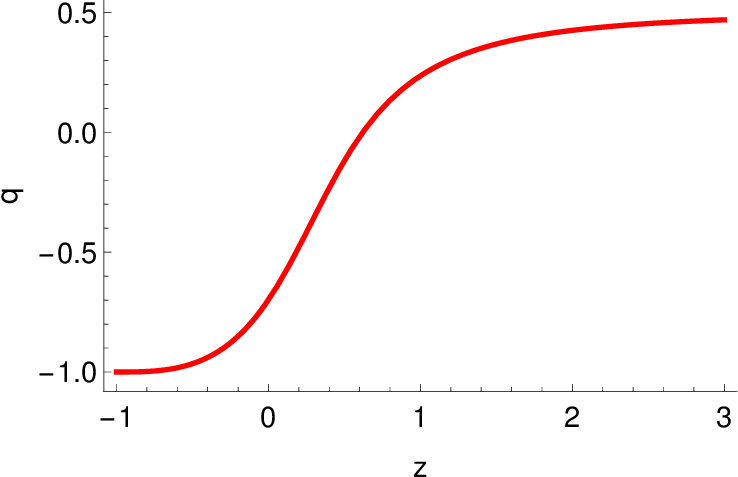} 
\caption{\it{{\bf{Upper   graph}}: 
Evolution of the  effective dark sector.  {\bf{ Upper graph}}: Evolution of the
energy density parameter $\Omega_{de}$} (blue-solid) and the
matter density parameter $\Omega_{m}$ (black-dashed) respectively, as a 
function of the redshift $z$, for the $f(Q)$ gravity scenario, in the case of $\Lambda \neq 0$.
    {\bf{ Middle graph}}: 
Evolution of the effective dark energy equation-of-state 
parameter $w_{de}$. {\bf{ Lower graph}}: The 
deceleration parameter $q$. In all graphs the parameter $\lambda=0.4$ in units of $H_{0,}$, and we have impose $ \Omega_{m0} \approx0.31$ according to observations. }
\label{OmegasL}
\end{figure}

Furthermore, the solutions of the dark energy density parameter $\Omega_{de}$ are still given by (\ref{omegade}) but with
\begin{eqnarray}
\nonumber &&
\!\!\!\!\!\!\!\!\!\!\!\!\!\!\!\!\!\!
\mathcal{A}=\frac{\lambda}{\Omega^{2}_{m0}(1+z)^{6}}, \\ \nonumber
&& \!\!\!\!\!\!\!\!\!\!\!\!\!\!\!\!\!\!\!\!
\mathcal{B}=144\lambda^{2}+2\lambda^{3}+54\left [\left (\frac{\lambda}{\mathcal{A}}\right )^{1/2}+\lambda +\frac{\Lambda}{3\Omega_{m0}(1+z)^{3}}\right ]^{2}, \\ \nonumber
&& \!\!\!\!\!\!\!\!\!\!\!\!\!\!\!\!\!\!\!\!
\mathcal{C}=\left [-4\left (24\lambda +\lambda^{2}\right )+\mathcal{B}^{2} \right ]^{1/2}.
\end{eqnarray}
In the upper graph of Fig. \ref{OmegasL} we present the evolution
of the energy densities $\Omega_{de}$ and $\Omega_{m}=1-\Omega_{de}$, in the middle graph the evolution of the equation-of-state parameter and in the lower graph the evolution of the deceleration parameter $q$. In all graphs we have $\lambda=0.4$ and $\Omega_{m0} \approx 0.31$ in agreement with the Planck results \cite{Planck:2018vyg}. We observe that with the same parameters, even an explicit cosmological constant is present, the cosmological behavior of the scenario at hand is approximatelly very similar with the case $\Lambda =0$, acquiring the usual thermal history of the universe, $w_{de}$ lies in the phantom regime and tends approximatelly to {-1} in asymptotically large times and the transition from deceleration to acceleration to happen at $z_{tr}\approx 0.62$. 
\begin{figure}[!h]
\centering
\includegraphics[width=7cm]{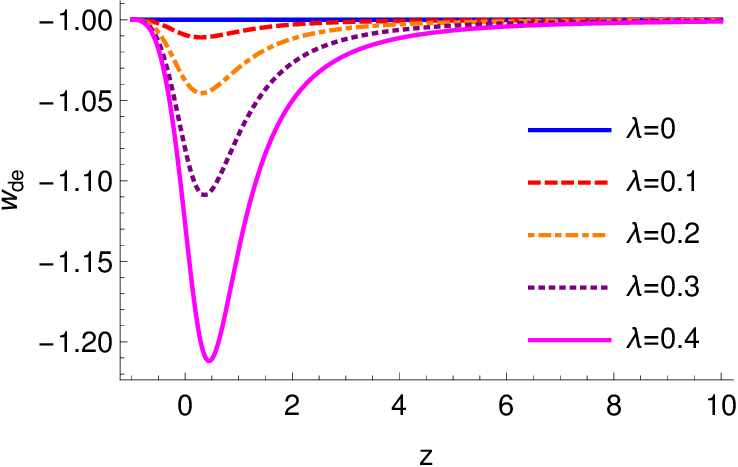}
\caption{\it{The evolution of the dark-energy equation-of-state parameter $w_{de}$ as a 
function of the redshift parameter $z$, for various values 
of the dimensionless parameter $\lambda $ in units of $H_0$. For each value of 
$\lambda$ we  choose  $\Lambda$ according to
(\ref{relL})  in order to obtain $\Omega_{m}(z=0)=\Omega_{m0}\approx0.31$ at present, 
and acquire the usual thermal history of the universe, similar to the upper graph of  
Fig.  \ref{OmegasL}.}}
\label{figwmultiL}
\end{figure}

Moreover, in Fig. \ref{figwmultiL} we depict the effect of the parameter $\lambda$ on $w_{de}$ in the case where an explicit cosmological constant is present. At high redishifts slightly lies in the phantom regime asymptotically close to $-1$, and near the present times deviates from $\Lambda$CDM cosmology, resulting always in dark energy dominated de-Sitter phase. Thus, we can see that the role of the parameter $\lambda$ has very similar effect in the two cases, \ie in the absence and in the presence of a cosmological constant.

In conclusion, $f(Q)$ gravity with the form (\ref{qform}), is very efficient in  
mimicking the cosmological constant, providing very interesting cosmological behavior.

\section{Conclusions} \label{concl}
 Motivated by the exciting features of the extended Symmetric Teleparallel Equivalent of General Relativity, namely the $f(Q)$ gravity, in this work we studied the cosmological evolution of the universe, for the two cases of the absence and presence of a cosmological constant, adopting a recent proposed generalized form of the $f(Q)$ function (\ref{qform}), which contain a single dimensionless parameter $\lambda$.

In particular, assuming dust matter, we extracted analytical expressions for the dark sector, namely for the dark energy density parameters, the equation-of-state parameter and for deceleration parameter, which quantified by the sole parameter $\lambda$.

In the case where an explicit cosmological constant is absent, these solutions show that the universe exhibits the usual thermal history, with the sequence of matter and dark-energy eras and the onset of acceleration at around $z\approx 0.62$ in agreement with observations. According to the value of $\lambda$, the dark-energy equation-of-
state parameter exhibits a very interesting behavior, slightly lying in the phantom regime at high redshifts, while it deviates at present times, and finally becomes around $-1$ in the future, resulting in a dark energy dominated, de-Sitter phase. Moreover, for this case we calculated the age of the universe, which coincides with the value corresponding to $\Lambda$CDM scenario and lastly, we confronted the scenario at hand with Supernovae type Ia data, in which case the aggrement is excellent.

Furthermore, for completeness, we consider an explicit cosmological constant, in order to compare the cosmological behavior of the scenario at hand in two cases. In this case,  
according to the value of $\lambda$, the dark energy equation-of-state parameter presents a quite similar behavior as in the case of the absence of a cosmological constant, deviating from the cosmological constant value at present time, while lying always in the phantom regime. Additionally, at asymptotic late times
it stabilizes in the cosmological constant value $-1$, resulting in a dark-energy dominated, de Sitter phase. From the cosmological behiavior of the scenario at hand in two cases, we saw that $f(Q)$ gravity with the form (\ref{qform}) provides a very interesting cosmological behavior, mimicking the cosmological constant in a very efficient way, and this is an advantage of the model.

In conclusion, modified gravity with non-metricity scalar Q, namely $f(Q)$ gravity, can provide exciting and new features in the study of the universe. Definitely, before
one considers it as a successful candidate for the description of dark energy, there are
necessary investigations that should be performed. In particular, one should confront the model with observational data from Supernova type Ia (SNIa), 
Baryon Acoustic Oscillation (BAO), Cosmic Microwave Background (CMB), weak lensing, full LSS spectrum, and Hubble parameter observations (OHD),
and extract constraints on the model parameter $\lambda$. Additionally, one should analyze in detail the phase-space behavior, in order to study the global dynamics and the asymptotic, late-time evolution of the model. The model at hand can be considered as a very promising alternative to the concordance model, and it would be both interesting and necessary, to study even more of its capabilitites. Such an investigation will be performed in a forthcoming publication.

\section{Aknowlegments}
The author is grateful to Emmanuel N. Saridakis, Fotios K. Anagnostopoulos and Smaragda Lola for helpful discussion and suggestions on the original manuscript. Special acknowledgement is due to the networking support by the COST Action CA18108 “Quantum Gravity Phenomenology in the Multimessenger Approach (QG-MM)”

\bibliographystyle{unsrt}
	\bibliography{fqrefs}

\end{document}